\documentclass[11pt,fleqn]{article} 
\title{Note on the Zero-Energy-Limit Solution for \\
the Modified Gross-Pitaevskii Equation}
\author{A. Kwang-Hua Chu} 
\date{Department of Physics, Xinjiang University,
Urumqi 830046, PR China}
\begin{document} 
\maketitle 
\begin{abstract} 
The modified Gross-Pitaevskii equation was derived and solved to
obtain the 1D solution in the zero-energy limit. This stationary
solution could account for the dominated contributions due to the
kinetic effect as well as the chemical potential in inhomogeneous
Bose gases.

\vspace{2mm}

\noindent %
PACS : \hspace*{2mm} 03.75.Fi, 03.65.Nk 
\end{abstract}
\doublerulesep=6mm 
\baselineskip=6mm 
\bibliographystyle{plain}
Studies of collision phenomena in rather cold gases, e.g., dilute
Bose gases, have recently attracted many researchers' attention
[1-2]. One relevant research interest is about the solution of the
appropriate and modified Gross-Pitaevskii equations for different
dimensions [3-4]. New possibilities for observation of macroscopic
quantum phenomena arises because of the recent realization of
Bose-Einstein condensation in atomic gases [1-2]. There are two
important features of the system - weak interaction and
significant spatial inhomogeneity. Because of this inhomogeneity a
non-trivial {\it zeroth-order} theory exists, compared to the {\it
first-order} Bogoliubov theory. This theory is based on the
mean-field Gross-Pitaevskii equation for the condensate $\psi$
-function. The equation is classical in its essence but contains
the ($h/2p$) constant explicitly. Phenomena such as collective
modes, interference, tunneling, Josephson-like current and
quantized vortex lines can be described using this equation. The
study of deviations from the zeroth-order theory arising from
zero-point and thermal fluctuations is also of great interest
[5-7]. Thermal fluctuations are described by elementary
excitations which define the thermodynamic behaviour of the system
and result in Landau-type damping of collective modes.
\newline As a preliminary attempt, followling the mean-field approximate formulation
in [3], in this letter, we plan to investigate the 1D solution for
the modified Gross-Pitaevski equation in the zero-energy limit.
This presentation will give more clues to the studies of the
quantum non-equilibrium thermodynamics in inhomogeneous (dilute)
Bose gases and the possible appearance of the kinetic mechanism
before and/or after Bose-Einstein condensation which is directly
linked to the particles (number) density and their energy states
or chemical potentials.
\newline The generalization of the Bogoliubov prescription [8] for
the $\psi$-operator to the case of a spatially nonuniform system
is
\begin{equation}
\hat{\psi} ({\bf r},t) \approx \psi_0 ({\bf r},t) + \hat{\phi}
({\bf r},t)
\end{equation}
where $\psi_0$ is the condensate wave function. This is an
expression of the second quantization ($\psi$-operator) for atoms
as $n_0=N_0/V=|\psi_0|^2$ and $\phi \ll \sqrt{n_0}$, $N_0$ is the
number of atoms in the condensate. To neglect $\hat{\phi}$ means
neglecting all correlations and this is a poor approximation when
distances between particles are of the order of the effective
radius $r_e$ of the atom-atom interaction. To overcome this
problem the assumption that the atomic gas is dilute : $n\, r_e^3
\ll 1$ should be made [1-3] ($n=N/V$, $N$ is the number of atoms
confined in $V$). We can use the procedure of the quantum virial
expansion to calculate the energy of the system. We have the form
\begin{equation}
E=E_0 + \frac{g}{2} \int n^2 ({\bf r}) d{\bf r},
\end{equation}
for the energy of slow particles, where $E_0$ is the energy of the
gas without interaction, $n({\bf r})$ os the density of the gas,
$g=4\pi a\, \hbar^2/m$, $a$ is the s-wave scattering length and
$m$ is the particle's mass [3]. After taking into account the
correlations (in above equation so that we can neglect
$\hat{\phi}$) and considering $E$ as an {\it effective
Hamiltonian} we then have the celebrated Gross-Pitaevskii equation
\begin{equation}
i \hbar \frac{\partial}{\partial t} \psi_0 ({\bf r},t)= [
-\frac{\hbar^2 \nabla^2}{2m} +V_{ext} ({\bf r}) +g |\psi_0 ({\bf
r},t)|^2 ] \psi_0 ({\bf r},t)
\end{equation}
which describe the dynamics of a non-uniform non-ideal Bose gas at
$T=0$. Here, $V_{ext} ({\bf r})$ is the confining potential. If
the gas is in its ground state, the time dependence of $\psi_0$ is
given by $\psi_0 \sim \exp(-i \mu t/\hbar)$, where $\mu$ is the
chemical potential of the gas [1-3]. We thus obtain the form
\begin{equation}
[ -\frac{\hbar^2 \nabla^2}{2m} +V_{ext} ({\bf r}) +g |\psi_0 ({\bf
r},t)|^2 -\mu] \psi_0 ({\bf r},t)=0
\end{equation}
which could be stationary once $t$ is fixed or selected. We
noticed that an equation of above form has been considered before
in connection with the theory of superfluidity of liquid helium
close to the $\lambda$-point [9]. \newline To investigate our
interest here, we shall consider the 1D solution of equation (4)
for the case of zero-energy limit. Firstly we consider a 3D Bose
gas confined tightly in one dimension and weakly in the remaining
two dimensions on a length scale $l_t$ (=$\sqrt{\hbar/2 m\omega}$
for a harmonic trap of angular frequency of $\omega$). A collision
between two condensate particles will typically occur over the
characteristic length scale $l_{col}$ once $l_t$ is much larger
than $l_{col}$ (thus we can use a local density approximation).
\newline
We now model the pair wavefunction of two atoms in the medium by
that a single particle with the reduced mass moving in a potential
which consists of a circularly symmetric box of radius $R$ and a
hard sphere of radius $R_a$ located in the centre of the box.
Following the reasoning in the derivation of equation (1), we
introduce a similar bias or ghost-effect for $\hat{\phi}$ : $\Psi$
which can be relevant to certain critical or kinetic
(non-equilibrium) effect (or configurational dissipation)
[1-2,5,7,10-11] so that $\mu \,\Psi =\varepsilon$ in the
zero-energy, zero-momentum limit. It is presumed that
$\varepsilon$ still reaches zero in the homogeneous limit. The
problem for a 2D $\psi (r,\theta)$ becomes, after referencing to
the bias $\Psi$,
\begin{equation}
[\frac{\partial^2}{\partial r^2}+\frac{\partial}{r\partial r}+
\frac{\partial^2}{\partial \theta^2}] \psi =-\varepsilon
\end{equation}
and, in fact, for a circularly symmetric $\psi (r)$,
\begin{equation}
[\frac{d^2}{d r^2}+\frac{d}{r\, dr}] \psi =-\varepsilon
\end{equation}
with the boundary conditions : the wave function vanishes on the
inner radius ($\psi =0$ as $r=R_a$), and reaches an asymptotic
value at the edge of the box ($\psi\rightarrow \Pi$ as $r=R$).
\newline We can obtain the solution
\begin{equation}
\psi (r)=\frac{\varepsilon}{4}[R^2-r^2
+(R^2-R_a^2)\frac{\ln(r/R)}{\ln(R/R_a)}]+\Pi
\frac{\ln(r/R_a)}{\ln(R/R_a)},
\end{equation}
where $R_a \le r \le R$. The extra energy caused by the curvature
of this wave function resulting from the scattering potential is
\begin{displaymath}
\Delta E = \frac{\hbar^2}{2m} \int_0^{2\pi}\int_{R_a}^R |\nabla
\psi (r)|^2 r dr d\theta=\pi \frac{\hbar^2}{m}\{\varepsilon
[\frac{R^3-R_a^3}{12}+\varepsilon
\frac{R^2-R_a^2}{\ln(R/R_a)}(\frac{1}{16}-\frac{R-R_a}{4})]+
\end{displaymath}
\begin{equation}
\frac{\Pi}{\ln(R/R_a)} [\Pi +\varepsilon
(R-R_a)(\frac{R-R_a}{2}-1)]\}.
\end{equation}
Note that, this energy depends upon the size of the box $R$, which
is indeed the length scale relevant for the scattering of two
particles in two dimensions [1-3]. The scattering of two particles
in a many-body system should obviously not depend on the size of
the system as a whole when $R$ becomes large, and so we must
interpret $R$ as the physical relevant length scale $l_{col}$. The
appropriate length scale over which a many-body wavefunction
changes is the healing length $l_h$, given in homogeneous Bose
condensed systems by $l_h
=\hbar/\sqrt{2m\,g_{2D}\,n_0}=\hbar/\sqrt{2m \mu}$ [1-3], and so
it is this which must be used in equation (8). We recall that $
|\Pi|^2$ corresponds to the condensate density $n_0$ and the
homogeneous limit for a pair interaction strength can still be
recovered from above equation. \newline If we  plot $\psi (r)$
w.r.t $r$ (in terms of units of $l_t$) for different $\varepsilon$
(0.0005, 0.001, 0.005, 0.01) and the same $\Pi$ (0.01) into a
figure then the subsequent presentation shows the significant
effect of the kinetic part (say, $\Pi$) due to $\varepsilon$. The
effect of boundary conditions, like $\Pi$ is minor. From the
definition of $\varepsilon=\mu \Psi$, we can understand that the
contribution of the chemical potential for $\psi$ of the
inhomogeneous gases is indeed dominated, too. We shall investigate
more complicated problems [12-13] by the same approach in the
future.


\begin{thebibliography}{99} 
\bibitem{RMP:2001} Leggett AJ 2001 {\it Rev. Mod. Phys.} {\bf 73} 307.
\bibitem{RMP:1999} Dalfovo F, Giorgini S, Pitaevskii LP, and
Stringari S 1999 {\it Rev. Mod. Phys.} {\bf 71} 463.
\bibitem{P:1999} Pitaevskii LP 1999 {\it Int. J. Mod. Phys. B}
{\bf 13} 427.
\bibitem{Chu:2001} Chu A K-H 2002 Preprint.
\bibitem{Salasnich:2001} Salasnich L 2001 {\it Int. J. Mod. Phys. B}
{\bf 15} 1253.
\bibitem{BEC:1956} Penrose O and Onsager L 1956 {\it Phys. Rev.} {\bf
104} 576.
\bibitem{T:Bose} Huang KS 1999 {\it Phys. Rev. Lett.} {\bf 83}
3770.
\bibitem{Bogoliu:1947} Bogoliubov NN 1947 {\it J. Phys. USSR} {\bf
11} 23.
\bibitem{P:G} Ginzburg VL and LP Pitaevskii 1958 {\it Sov. Phys.
JETP} {\bf 7} 858.
\bibitem{Kagan:1992} Kagan YuM, Svistunov, and Shlyapnikov GV 1992
{\it Sov. Phys. JETP} {\bf 74} 279.
\bibitem{1989:Bose} Snoke DW and Wolfe JP 1989 {\it Phys. Rev. B}
{\bf 39} 4030.
\bibitem{GP:N} Adhikari SK and Muruganandam P 2002 {\it J. Phys. B
At. Mol. Opt. Phys.} {\bf 35} 2831.
\bibitem{Chu:PostDr} Chu K-H 2001 Post-Dr. Report (in English; Peking University, Beijing).
\end{thebibliography}
\end{document}